\documentstyle[12pt,preprint]{aastex}

\def\boo{Bo{\"o}tes }

\lefthead{}
\righthead{}

\begin{document}
\title{RR Lyrae Stars in the \boo dSph}

\author{M. H. Siegel}
\affil{University of Texas, McDonald Observatory}
\authoraddr{1 University Station, C1402, Austin TX, 78712\\
e:mail: siegel@astro.as.utexas.edu}

\begin{abstract}
We present a catalog of 15 RR Lyrae variable stars in the recently discovered \boo dSph
galaxy -- the most metal-poor simple stellar population with measured RR Lyrae stars.  
The pulsational properties of the RR Lyrae conform closely to period-abundance trends extrapolated from 
more metal-rich populations and we estimate the distance of \boo to be $(m-M)_0=18.96\pm0.12$.
The average period (0.69 days), the ratio of type c to type ab 
pulsators (0.53) and the RRab period shift (-0.07) indicate an Oosterhoff II classification for
Bo{\"o}tes, a marked contrast to the other
dSph galaxies, which are Oosterhoff intermediate.
This supports the contention that the Oosterhoff dichotomy is a continuum --
that RR Lyrae properties, to first order, vary smoothly with abundance.
The dSph galaxies are not distinct from the Galactic globular clusters, but bridge
the Oosterhoff gap.  The absence of
any anomalous Cepheids in \boo could indicate the lack of an intermediate age population.
\end{abstract}
\keywords{stars:variables:RR Lyrae -- galaxies: dwarf -- galaxies: individual (Bo{\"o}tes)}

\section{Introduction}
Belokurov et al. (2006, hereafter B06) have reported the discovery of a new dSph galaxy 
in the constellation of Bo{\"o}tes.
This galaxy, claimed to be faintest known Local Group galaxy, is very sparse and apparently metal-poor, 
with a color-magnitude diagram consistent with an abundance near 
$[Fe/H]\sim2.3$.  Munoz et al. (2006, hereafter M06) have presented spectra of \boo stars,
revealing the dSph to be 
extremely metal-poor ($[Fe/H]\sim-2.5$) with a high apparent M/L ratio ($\sim130-680$).

Studies of variable stars in dSph galaxies (Siegel \& Majewski 2000, hereafter SM00, and references
therein) have found rich populations of RR Lyrae variable stars as well as anomalous Cepheids.
The variable stars are useful not only for constraining the distance modulus to the dwarfs 
but for confirming gross properties of the stellar populations.  The presence of anomalous
Cepheids and RR Lyrae stars indicate intermediate and old stellar populations, respectively, and 
the pulsational properties of the latter are heavily dependent upon the abundance --
and perhaps evolution -- of the horizontal branch (HB) stars.  Inversely, the RR Lyrae in dSph
galaxies may probe regions of parameter space unoccupied by Galactic RR Lyrae.  \boo is a perfect
example of this -- it appears to be one of the most metal-poor simple stellar populations near the
Galaxy.  In this Letter, we present
a survey of variable stars in Bo{\"o}tes, identify 15 RR Lyrae variable stars and show
that their pulsational properties are consistent with trends extrapolated from more
metal-rich populations.  We also comment on the impact
this study has on the Oosterhoff ``dichotomy''.

\section{Observations and Reduction}

We observed the \boo dSph with the 0.8-meter telescope at McDonald Observatory on UT 
April 30 to May 3 2006 and June 23-26 2006.  Images were obtained with the Prime Focus Corrector
(PFC) in the $B$, Washington $M$, $T_2 (=I)$ and narrow-band $DDO51$ filters.  The PFC 
has a field of view of 46\farcm0 and although this should presumably be large enough to 
observe the entirety of Bo{\"o}tes, a slight offset in the 0.8-m setting circles displaced
our pointing slightly southwest of the dSph center.  \boo subtends a large angle
on the sky (M06 estimate $r_h\sim$13\farcm) and there is some indication that it 
extends off the eastern end of our survey field.

Data were reduced with the IRAF CCDPROC package.  Although the 0.8-m telescope 
is unguided, the tracking is stable enough -- and the pixel scale coarse enough -- to allow 600s
integrations without significant image ellipticity.  Only the first two nights of the April run
were photometric.

The data were photometered using the DAOPHOT and ALLFRAME packages (Stetson 1987, 1994).
Despite 
the coarse pixel scale of the PFC (1\farcs35 per pixel), we derived excellent PSFs with 
precise photometry down to Bo{\"o}tes' HB.  Photometric errors are approximately
$\sigma_B=0.08$ at $B=20$ for individual images and $\sigma_B=.01$ at $B=20$ in the combined image
created for ALLFRAME.

Standard stars were measured using multi-aperture photometry and DAOGROW
(Stetson, 1990) to extract total magnitudes.  The total magnitudes were then 
calibrated to the standards of Landolt (1992) and Geisler (1990,1996)
using the matrix inversion methods outlined in Siegel et al. (2002).  Individual images
were calibrated using the iterative technique described in Siegel et al. with apertures
corrected to the total magnitudes of the PSF stars as calculated by DAOGROW.
The $B$ zero point of the photometric frames showed variation consistent 
with that of the standard stars (.01 mag).

RR Lyrae stars vary in color by 0.2-0.4 mag (Nemec 2004), 
which could induce offsets in
the relative photometry.  Although we took four $I$
observations from which stellar colors could be measured, this is inadequate
for precise epochal colors for each CCD frame.  We therefore transformed 
the photometry of the RR Lyrae
stars using the average RR Lyrae $<B>-<I>$ color. Given that the $B-I$ color term 
is small (-.059 mag), this reduces the color-induced photometric scatter to $<1-2$\%.

Accurate stellar positions were estimated from the IRAF task TFINDER and the NOMAD
astrometric catalog (Zacharias et al. 2004).  The astrometry has a precision of 0\farcs2
in each coordinate.

\section{Light Curve Fitting}

Light curves are based on 58 $B$ images obtained with the PFC.  A 
table of Julian dates and $B$ magnitudes 
for all our RR Lyrae stars is available electronically through the ApJ.

To identify variable stars in our data, we used the variability index produced by 
DAOMASTER -- the ratio of scatter to observational error.  We selected stars with
variability greater than 2.75 as potential variables.
Twenty stars in our sample showed this level of variability.  Figure 1 shows the 
color-magnitude diagram of \boo with the variable stars marked.  

Fifteen of 
the variables are on the \boo HB and all of these are RR Lyrae stars (\S4).
The bright star near $B-I\sim1.5$ 
shows 0.1 mag variability with no consistent periodicity on the 1-2 day timescale.  Its
$M-DDO51$ color is precisely along the field dwarf locus.
All three red variables have $M-DDO51$ colors consistent with stars near the tip of 
the red giant branch (TRGB).  One of these has an apparent magnitude ($I=14.7$) near
the \boo TRGB ($I=15.2$; Bellazzini et al. 2004).  The red stars show variations of 0.1-0.2 mag, 
but manifested {\it between} the two observing runs, not over either individual 
run.  These may be long-period variables near the TRGB.
The remaining bright variable is located
just above the \boo HB -- a position normally associated with anomalous Cepheid variables.  However,
this star shows no periodicity on the scale of 1-2 days, only a 0.2 magnitude difference
between the observing runs.  It could be a foreground quasar or another long-period variable.

The absence of any anomalous Cepheids (AC) could indicate that \boo lacks an intermediate-age population
(see Nemec et al. 1988; Mateo et al. 1995).  However, \boo may have too few stars
to produce AC.  Our study of Leo II (SM00) identified only
four AC against 148+ RR Lyrae stars.

For the RR Lyrae candidates, periods were fit using methods outlined in SM00.  We used
Stetson's (1996) modified version of the Lafler-Kinman index (1965) to identify potential periods and used
a Levenberg-Marquardt algorithm to fit the RR Lyrae templates of Layden (1998) to 
each trial period.  With each star, a clear minimum $\chi^2$ was found\footnote{Star V3 has
two $\chi^2$ minima at .32 and .48 days.  We did not capture the star during its
rise and both periods fit the descent of the light curve -- although with a much larger amplitude
for the .48 day fit (both were RRc templates).  We selected the 0.32 day fit
because the period and amplitude would be more consistent with the other \boo RRc stars as well
as those in other dSph galaxies and globular clusters.}.  We should expect 1-3 of Bo{\"o}tes'
RR Lyrae to demonstrate the Blazhko (1907) effect but our data are not extensive enough to detect
second-order variation.  None of our stars exhibit double-mode pulsation
(Cox et al. 1980; Sandage et al. 1981; Cox et al. 1983; Nemec 1985a).

Periods were checked by comparison to the phase dispersion minimization (Stellingwerf 1978)
program in IRAF and the STDLC template-fitting program of Layden et al. (1999) and 
Layden \& Sarajedini (2000).  Both fit periods within measurement uncertainties of ours.

\section{\boo RR Lyrae Stars}

Table I lists ID, coordinates, period, amplitude and Bailey (1902)
type (pulsation mode) for the 15 RR Lyrae stars identified in Bo{\"o}tes.  
We also list intensity-weighted mean
magnitudes ($m_B$) calculated by integrating the Layden templates in .02 
phase increments at the fit amplitude and period.  Light curves of all our variables are shown
in figure 2.

Figure 3 shows the period-amplitude distribution of the RR Lyrae stars in Bo{\"o}tes.  We
note that they follow a period-amplitude relationship similar to that seen in other
globular clusters and dSph galaxies, but offset to longer periods -- a result consistent
with a very low abundance for Bo{\"o}tes.  The c variables are of nearly constant
amplitude but show a hint of the parabola shape predicted by Bono et al. (1997).

\subsection{\boo Stellar Populations}

Sandage (1993, S93) demonstrated that the pulsational properties of RR Lyrae track
the abundance of the parent population.  This provides a check on abundance
that is independent of photometric zero point, reddening or assumptions about
standard candles.  \boo would represent the lowest-metallicity population
for which these relations have been tested.

The most reliable diagnostic of S93 is the average period of the RRab stars, which is
updated in Sandage (2006) to:

\begin{center}
log$<P_{ab}>$ = -0.098 [Fe/H] - 0.416
\end{center}

The RRab variables in \boo have an average period of $.691\pm.089$ days, consistent
with an abundance of [Fe/H]=-2.6 on the Zinn-West (1984) scale.

The shortest and longest RRab period depends on the location of the
blue and red edges, respectively, of the instability strip.  S93 showed that these values
also depend on the abundance of the parent population.  We use the formula
for the the shortest period from Sandage (2006) and for the longest period from S93:

\begin{center}
log($P_{ab}$) = -0.452 + 0.033 [Fe/H]$^2$
\end{center}

\begin{center}
log($P_{ab}$) = -0.09 [Fe/H] - 0.280
\end{center}

Our shortest RRab has a period of 0.576 days, 
consistent with an abundance of [Fe/H]=-2.5 on the Zinn-West scale.
Our longest period variable (.859 days) would be consistent with an abundance of 
[Fe/H]=-2.4
on the Butler-Blanco scale (Butler 1975; Blanco 1992), which we correct
to a Zinn-West abundance of [Fe/H]=-2.6.

Finally, the average RRc period tracks abundance by the S93 equation:

\begin{center}
log$<P_{c}>$ = -0.119 [Fe/H] - 0.670,
\end{center}

\noindent which, given the average \boo RRc period of $.364\pm.044$ days, indicates an abundance
of [Fe/H]=-2.0.  This is more metal-rich than the other measures.  We would
expect the RRc stars to have an average period of 0.42 days given the S93 formula.
If star V3 were to be evaluated at the 0.48 day degeneracy, this would only increase
the derived abundance to $[Fe/H]\sim-2.1$.  A degeneracy in an RRab star and/or
several undiscovered RRc with periods near 0.5 days would be required to move the
average period to longer values.  The average RRC period is the only measure inconsistent 
with the measured abundance of Bo{\"o}tes.  This may indicate non-linearity in the $[Fe/H]-<P_c>$
relation, similar to the slight non-linearity identified in the relationship between 
abundance and shortest RRab period.

It should be noted that \boo subtends a large solid angle -- its variables are distributed
across the PFC field with 14/15 on the eastern half.  A few RR Lyrae 
stars may have been missed in our survey.
Given the small number of RR Lyrae stars, a missed variable with an extremely long or
short period could alter our results, shifting the average period or providing an RRab
with a shorter or longer period than the ones we have measured.

Nevertheless, our pulsational diagnostics {\it are} consistent with the very low \boo abundance
indicated by B06 and
M06.  This demonstrates that the connection between RR Lyrae pulsation properties and abundance
extends to the extremely metal-poor domain.  More variable star surveys of the growing number of
newly-discovered metal-poor dwarfs in the Local Group would test this assertion.

We noted above that the period-amplitude relation of Bo{\"o}tes' RR Lyrae stars is shifted
relative to M3.  Period shifts are well-established in cluster RR Lyrae
(Sandage 1981a; Sandage 1981b; Carney et al. 1992) and are quantitatively defined
by comparison to a reference variable star population, M3 being the usual template.
SM00 refine the period shift measure to:

\begin{center}
$\Delta log P = -[0.129 A_{B} + 0.112 + log P]$
\end{center}

The $\Delta log P$ values of \boo are plotted in Figure 4.  The right
ordinate has the $\Delta M_{bol}$ scale derived in SM00 by applying assumptions about period shift 
(notably constant mass) to the fundamental pulsation equation of van Albada \& Baker (1971).

The RR Lyrae stars in \boo have a mean period shift of -.07.  There is some
scatter in the period shifts which may indicate scatter in the fundamental properties
of Bo{\"o}tes' RR Lyrae -- either age, abundance or evolution.  The period shifts of Bo{\"o}tes' RR Lyrae are
consistent with bolometric luminosities several tenths of a magnitude brighter than
the RR Lyrae of M3, as expected for a very metal-poor population.  We tested this hypothesis
by comparing $m_B$ against $\Delta log P$. We found that
a linear correlation with a slope of 0.7 mag.  This would be consistent with the
variable stars of \boo being 0.05 mag brighter than those of M3, a luminosity difference
four times smaller than that predicted by the magnitude-abundance relations in the 
literature (see \S5).  This indicates either that the assumption of constant mass in
the SM00 derivation is erroneous or that there is too much scatter in the period shift for an
{\it internal} check on the $m_B-\Delta log P$ relation.

\subsection{\boo and the Oosterhoff Continuum}

Oosterhoff (1939) established that Galactic globular clusters can be divided
into two categories based on the average RRab period and RRab to RRc ratio 
(see review in Catelan 2005).  The dSph galaxies, LMC and Fornax globular clusters, 
and M31 have average $<P_{ab}>$ and $\frac{N_{RRc}}{N_{RRab}}$ measures intermediate between 
OoI and OoII (SM00;
Catelan 2005 and references therein), filling the Oosterhoff gap and providing a continuous
trend of pulsation properties with abundance (see Figure 6 in SM00).  
Nevertheless, some studies of RR Lyrae stars continue to differentiate
stars based on their Oosterhoff type.

It could be argued that the dSph RR Lyrae are different from the globular cluster
RR Lyrae by virtue of the second-parameter effect.  However, \boo breaks this
trend.  It is very metal-poor -- the most metal-poor object known to
have RR Lyrae stars.  
The average period of \boo RRab stars (0.69 days) and the ratio of RRc to RRab (0.53) place \boo
in the OoII category.  The Oosterhoff dichotomy also manifests in period shift -- both
cluster and field stars avoid $\Delta log P$ values between 
-0.01 and -0.05 (Suntzeff et al. 1991, marked in Figure 4).  The stars in \boo have OoII period shifts.

This means that the dSph galaxies
not only fill the Oosterhoff gap in the Peterson diagram, but {\it bridge} the
gap, providing a continuum of behavior from OoI to OoII.  This reinforces the contention
of Renzinini (1983), Castellani (1983) and S93 that the Oosterhoff gap only exists
because of blueward shift in the HBs of intermediate-abundance Galactic globular clusters.

Of course, as SM00 note, there are many parameters that affect RR Lyrae pulsational
properties besides abundance -- notably evolution from the zero-age HB
(Lee, Demarque \& Zinn 1990; Lee \& Carney 1999; Demarque et al. 2000).  Nevertheless, the presence of \boo in the 
OoII part of the Peterson diagram shows
that the dSph galaxies follow the same abundance-pulsation relations as the globular clusters.
It is clear that when {\it all} groups
of RR Lyrae stars are considered, metallicity is the ``first parameter" of RR Lyrae
pulsational properties.  The data require no fundamental difference between OoI and 
OoII RR Lyrae stars
because no dichotomy exists in the observational plane.

\section{Distance and $M_V$ of \boo}

RR Lyrae stars are excellent standard candles.
The average $m_B$ of Bo{\"o}tes' RR Lyrae variables is $B=19.81\pm.05.$\footnote
{The RRc variables are .05 mag brighter than the RRab variables.}  Sandage (2006) estimates that
the typical RR Lyrae at the abundance of \boo have a $(B-V)_0$ color of $0.331\pm.025$.  Piersimoni et
al. (2002) provide an empirical calibration of RRab color from period, $A_B$ and 
[Fe/H], from which we calculate $(B-V)_0=.332\pm.03$ for our stars.  
This places the RR Lyrae locus of \boo at $V=19.48\pm.06$.  An analysis of the literature
(Fernley et al. 1998; Gratton et al. 2004; Sandage 2006) indicates
that stars at $[Fe/H]=-2.5$ should have an absolute magnitude of $M_V=0.35\pm0.10$, resulting
in a \boo distance modulus of $(m-M)=19.13\pm0.12$.  B06 estimate an $A'_V$ extinction
value for \boo of 0.06 mag.  
We find,
from the maps of Schlegel et al. (1998), a reddening of $E_{B-V}=.056; A'_V=0.17$.  This produces
an absolute distance modulus of $(m-M)_0=18.96\pm.12$ for a distance of $62\pm4$ kpc.

Our survey reveals more variable stars in \boo
than would be expected from such a low luminosity dwarf ($M_V=-5.8$; B06).  By comparison, SM00 detect
ten times as many RR Lyrae in Leo II, which is over four magnitudes ($\sim80\times$) brighter than 
Bo{\"o}tes. 
Our preliminary examination of Ursa Major, which B06 claim is of similar
brightness to Bo{\"o}tes, indicates perhaps one RR Lyrae in the field.  In fact, \boo has more RR Lyrae
stars within its core radius (12) than Ursa Major has HB stars (9, from figure 1 of Willman et al. 2005).
This could indicate that \boo has a high frequency of RR Lyrae stars -- or it could indicate that B06
underestimate Bo{\"o}tes' $M_V$.  An absolute magnitude of $M_V\sim-7$ would be more consistent with
the number of RR Lyrae stars in Bo{\"o}tes.

\section{Conclusions}

We identify and fit periods
to 15 RR Lyrae variable stars in the \boo dSph galaxy.  While \boo subtends a large solid angle and may have RR Lyrae outside
our survey field, the average, shortest and longest RRab periods are consistent with the very low
metallicity derived by B06 and M06.
\boo is the most metal-poor object in which RR Lyrae
have been identified.  The distance modulus
of \boo is $(m-M)_0=18.96\pm0.12$.

\boo is a canonical OoII object, with long RRab periods, a large fraction of RRc variables and
large negative RRab period shifts.  In combination with the other dSph galaxies, it bridges the Oosterhoff
gap, confirming that the the gap is a selection effect peculiar to the 
Milky Way and not necessarily a fundamental aspect of RR Lyrae variable stars.

\acknowledgements

This research was supported by NSF grant AST-0306884.  Work on this program was performed at the Aspen
Center for Physics.

\newpage

\begin{center}
TABLE II. RR Lyrae Variable Properties\\
\begin{tabular}{c|c|c|c|c|c|c} \tableline\tableline
Var ID &   RA	      &  DEC	  & Period  & $A_{B}$ & $m_{B}$  & Bailey \\
       &   J2000.0    & J2000.0   & (Days)  &	      & 	 & Type   \\
\tableline
V1  & 13:59:29.36 & 14:10:43.6 & .3037714 &   0.628  &  19.78 & c \\   
V2  & 13:59:51.34 & 14:39:06.1 & .3102270 &   0.667  &  19.73 & c  \\
V3  & 14:00:26.87 & 14:35:33.3 & .3228953 &   0.638  &  19.79 & c   \\
V4  & 14:00:08.90 & 14:34:24.3 & .3832322 &   0.584  &  19.82 & c   \\
V5  & 14:00:21.56 & 14:37:29.0 & .3863158 &   0.566  &  19.75 & c   \\
V6  & 13:59:45.95 & 14:31:40.8 & .3918305 &   0.607  &  19.86 & c   \\
V7  & 13:59:49.37 & 14:10:05.6 & .4011623 &   0.742  &  19.76 & c   \\
V8  & 13:59:59.69 & 14:27:34.1 & .4145401 &   0.489  &  19.79 & c  \\
V9  & 13:59:47.28 & 14:27:56.4 & .5758855 &   1.279  &  19.84 & ab  \\
V10 & 14:00:25.76 & 14:33:08.6 & .6279676 &   1.325  &  19.78 & ab  \\
V11 & 13:58:04.38 & 14:13:19.3 & .6617310 &   1.201  &  19.82 & ab  \\
V12 & 13:59:56.00 & 14:34:55.1 & .6797488 &   0.544  &  19.90 & ab  \\
V13 & 13:59:06.36 & 14:19:00.1 & .7061108 &   0.819  &  19.77 & ab  \\
V14 & 13:59:25.75 & 14:23:45.4 & .7244797 &   0.858  &  19.89 & ab \\
V15 & 14:00:11.08 & 14:24:19.7 & .8586484 &   0.481  &  19.83 & ab \\
\tableline
\end{tabular}
\end{center}

\newpage

\figcaption[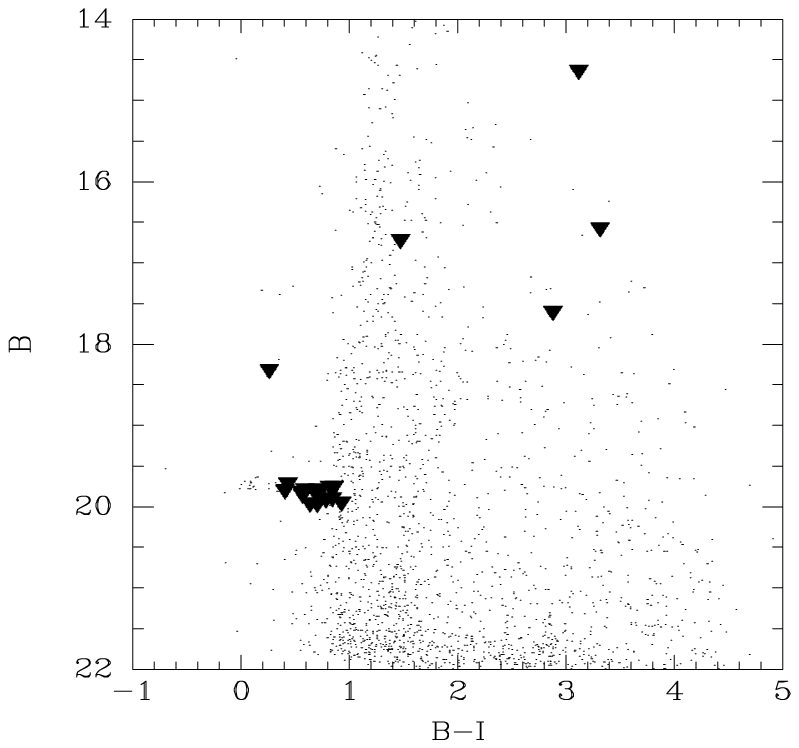]{$BI$ color-magnitude diagram of the \boo field.  Variable stars
are marked with triangles.}

\figcaption[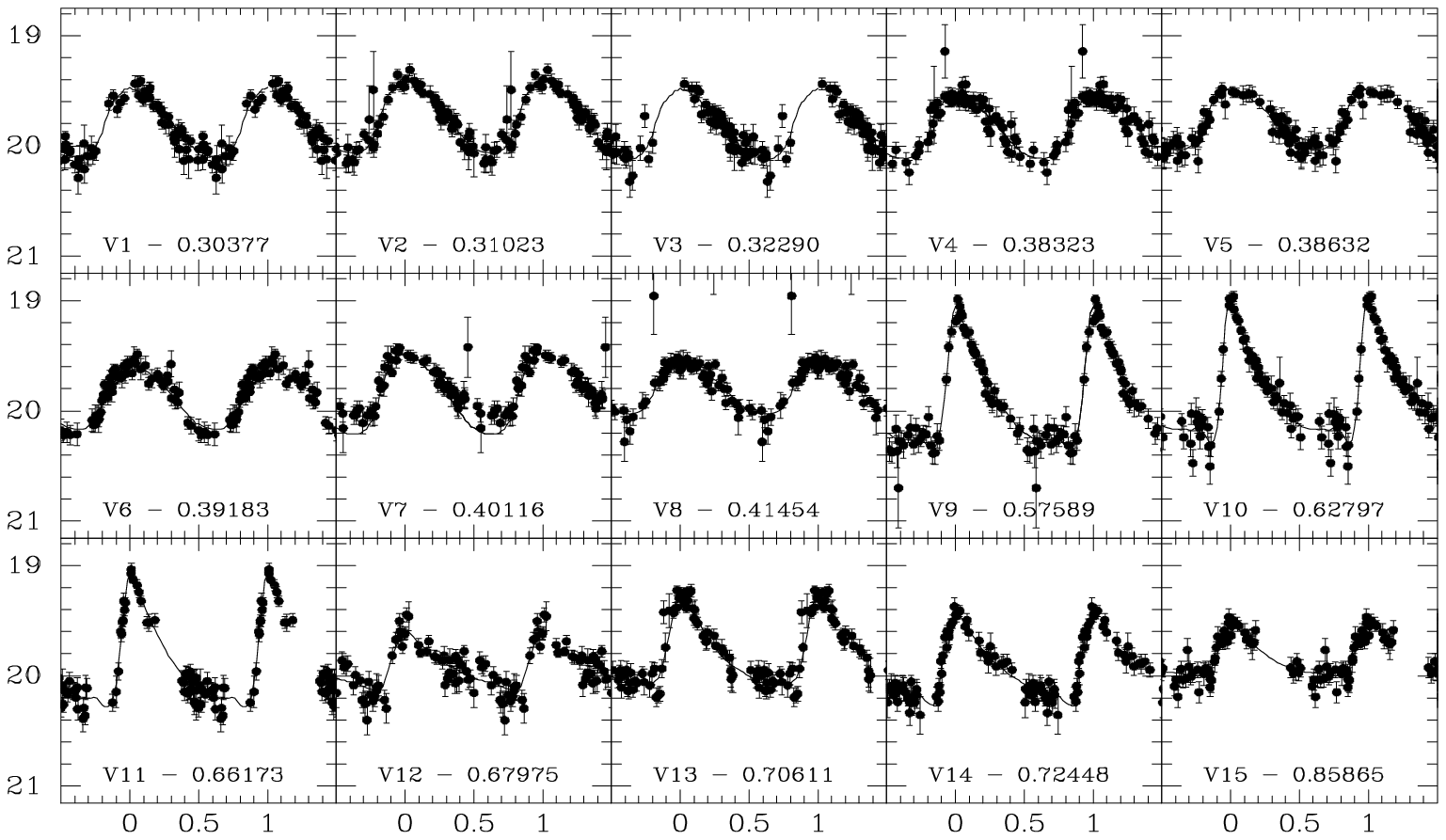]{Light curves for the \boo RR Lyrae, sorted by increasing period.
Template light curves are overlayed for comparison.}

\figcaption[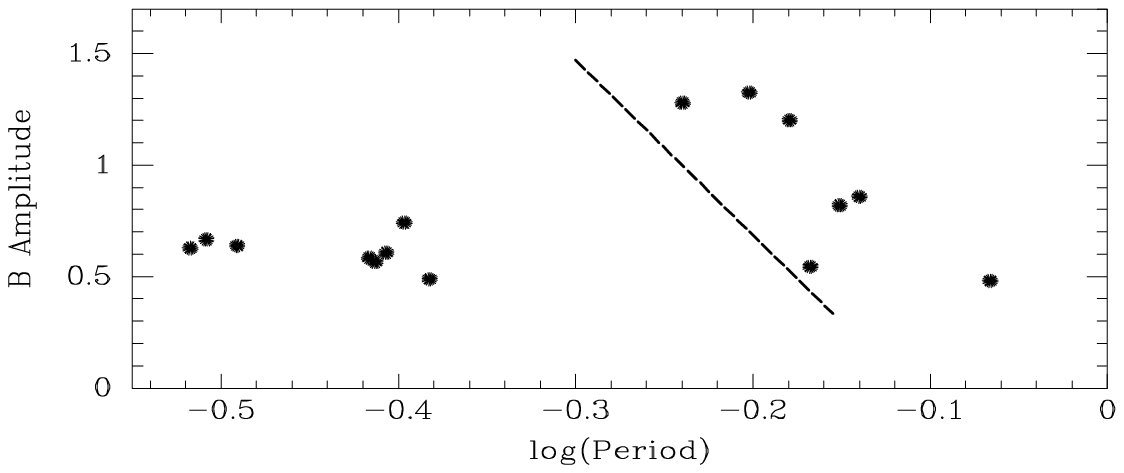]{Period-amplitude distribution of the RR Lyrae stars in Bo{\"o}tes.  The dashed
line marks the period-amplitude locus of M3, parameterized by SM00.}

\figcaption[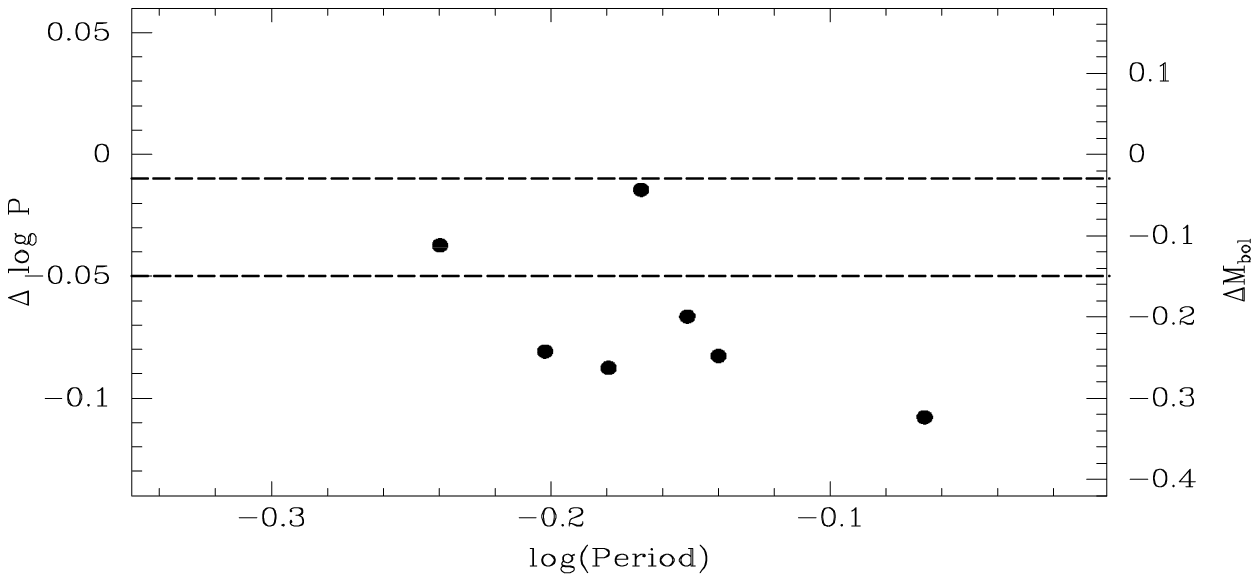]{The period shift ($\Delta log P$) of Bo{\"o}tes' RR Lyrae stars.  
The right
ordinate shows the corresponding bolometric magnitude shift, under the assumption of constant
mass.  The dashed lines mark the 
zone avoided by Galactic field and globular cluster stars.  This may be contrasted with a similar diagram
for Leo II shown in SM00 and Mateo et al. (1995).}

\newpage

\begin{figure}
\plotone{f1.eps}
\end{figure}
\begin{figure}
\plotone{f2.eps}
\end{figure}
\begin{figure}
\plotone{f3.eps}
\end{figure}
\begin{figure}
\plotone{f4.eps}
\end{figure}

\end{document}